\documentclass[a4paper]{jpconf}
\usepackage{graphicx}
\usepackage{amsmath} 
\usepackage[utf8]{inputenc}
\usepackage{times}
\usepackage{graphicx}
\usepackage{float}
\usepackage{subfig}
\begin{document}
\title{Significance of lower energy density region of neutron star and universalities among neutron star properties.}
\author{$^1$Wasif Husain}
\ead{wasif.husain@adelaide.edu.au}
\author{$^1$Anthony W. Thomas}
\ead{anthony.thomas@adelaide.edu.au}

\address{$^1$Department of Physics, The University of Adelaide, North Terrace, Adelaide 5005.}
\begin{abstract}
We have constructed and compared models of rotating neutron stars and strange stars, within the Hartle framework. The significance of the low energy density region and crust region inside the neutron star has been studied, along with how much the existence of strange matter above the energy density 300 MeV/fm$^3$ can affect the neutron star properties. We have confirmed several universalities among the neutron star properties such as, dimensionless moment of inertia  vs dimensionless quadrupole moment, dimensionless tidal deformability vs dimensionless moment of inertia and moment of inertia parameters vs R/2M.
\end{abstract}
\section{Introduction}
From the atmosphere to the core of neutron stars covers an energy density range from 10$^6$ gm/cm$^3$ to several times nuclear matter density. At  densities above nuclear matter density matter may undergo phase transitions and new exotic states may form. The existence of these states relies on the strong interaction and the structure of baryons. In 1960 [1] it was suggested that the core of neutron stars may have mesons (pions-the lightest meson). In 1980 [2]  the possibility of Bose condensation of Kaons at densities above three times nuclear matter density was suggested. Since then it has been addressed by many authors, for instance see Ref.[3]. As the energy density increases, baryons may  decompose into their constituent quarks. Invanenko and Kurdgelaidze [4] in 1965 showed that neutron stars core may be made of deconfined quark matter. Since 1990, such phase transitions have been considered by many authors [5][6].
\par The key ingredient to understand their interior is the nuclear equation of state. New born neutron stars are extremely hot but after  enough time they become cold and they have matter made of mainly neutrons partly protons, electrons and muons in $\beta$ equilibrium with respect to the weak interactions. Different approaches have been adopted over  time to model the neutron star under the constraints imposed by observed neutron star properties. We will compare properties of neutron stars based on EoSs that include either, nucleons only, nucleons and hyperons or strange quark matter at high energy density.
\section{Models of compact stars}
\subsection{Equation of state (EoS)} We are taking into consideration three EoSs. N-QMC700 and F-QMC700 (see Ref.[7]) which are based on QMC model [8]. N-QMC700 only allows nucleons while F-QMC700 includes hyperons. Both EoSs take nucleon structure into account. In both, N-QMC700 and F-QMC700, the mass of $\sigma$ meson [9] is considered $m_\sigma$ = 700 MeV . Below energy density $<$ 110 MeV/fm$^3$, the matter forms in the crust of the star, with inhomogeneities consisting of nucleons arranged on a lattice, as well as neutron and electron gases. In this density region the QMC Equation of State (EoS) needs to be matched [7] with the equations of state, showing the composition of matter at those densities. Baym-Pethick-Sutherland (BPS) EoS [10] is used in this work. For the sake of convenience the parametrized EoSs are given below. The parameterization of the EoSs of N-QMC700 and F-QMC700 is
\begin{equation}
P = \frac{N_1\epsilon^{p1}}{1+e^{(\epsilon - r)/a}}  + \frac{N_2\epsilon^{p2}}{1+e^{-(\epsilon - r)/a}} 
\end{equation}
which works well up to an energy density  1200 MeV/fm$^3$ and values for the constants $N_1$, $N_2$, $p_1$, $p_2$, '$r$' and '$a$' are given in Table 1. 
\begin{center}
	\begin{table}
		\begin{tabular}{|c|c|c|c|c|c|c|}
			\hline
			& $N_1$ & $p_1$ & $N_2$ & $p_2$ & $r$ & $a$ \\ [0.5ex] 
			\hline\hline
			N-QMC700 &   0  & 0 &0.008623& 1.548& 342.4 &184.4\\
			\hline
			F-QMC-700 &0.0000002.62& 3.197& 0.0251& 1.286& 522.1 & 113\\
			\hline
		\end{tabular}
		\caption{Parameters of EoS (1) from Ref.[7]}
	\end{table}
\end{center}
The results will be compared with the EoS [11] of strange stars which is based on the MIT bag model [12]. 
The EoS of a strange star matter (strange matter EoS or MIT EoS) is given [11] by
\begin{equation}
P = \frac{1}{3}(\epsilon - 4B),
\end{equation}
where B is the bag constant whose value is taken to be $10^{14}$gm/cm$^3$. This EoS suggests quark deconfinement at the higher energy densities.
\subsection{Structural equations}
The structural equation for a static, spherical neutron star involves solving the TOV [13][14] equation, derived from the general theory of relativity [15], for a particular equation of state (EoS). The line element for such a compact star is given by (G = c = 1)
\begin{equation}
ds\textsuperscript{2} = -e^{2\Phi(r)}  dt^{2} + e^{2\Lambda(r)} dr^{2} + r^{2} d\theta^{2} +r^{2}\sin^{2}\theta d\phi^{2}.
\end{equation} 
The TOV equation for the star is 
\begin{equation}
\begin{aligned}
\frac{dP}{dr} = - \frac{[\epsilon(r)+P(r)][4\pi r^3P(r)+m(r)]}{r^2(1-\frac{2m(r)}{r})}, \qquad m(r) = 4\pi\int_{0}^{r}dr.r^2 \epsilon(r).
\end{aligned}
\end{equation}
\paragraph{Rotational equation.}
For slowly rotating neutron stars, Hartle's [16] approach is adopted. Structural equations are solved under the general relativistic framework. Hartle's approach to get the solution for a perturbed Schwarzschild metric is based on the idea that,  as stars rotate their pressure, energy density and baryon number, all get perturbed. The perturbed line element of rotating, axially symmetric equilibrium configuration, is given by
\begin{equation}
ds^2 = -e^{2\nu(r,\theta,\Omega)}(dt)^2 + e^{2\psi(r,\theta,\Omega)}(d\phi - \omega(r,\theta,\Omega)dt)^2 + e^{2\mu(r,\theta,\Omega)}(d\theta)^2 + e^{2\lambda(r,\theta,\Omega)}(dr)^2 + \mathcal{O}(\Omega^3)
\end{equation}
where $\nu$, $\psi$, $\mu$ and $\lambda$ are the perturbed metric functions , $r$ and $\theta$ are polar coordinates and $\Omega$ is  the uniform rotational velocity of the star. $\omega$ is the rotating velocity of the local inertial frame of reference dragged along the direction of rotation of the neutron star. This dragged velocity also depends on the polar coordinates $r$, $\theta$. Since the velocity of local dragging of the inertial reference frame depends on the neutron star's mass concentration inside and outside, which varies with $\Omega$, $\omega$ is a function of $\Omega$. Relative angular velocity is denoted by, $\overline{\omega} (\Omega - \omega(r,\theta,\Omega) )$
It is this relative velocity ($\overline{\omega}$), which is of particular interest, when discussing the rotational flow of the fluid inside the neutron star.
\begin{equation}
\frac{d}{dr}(r^4j(r)\frac{d\overline{\omega}_{l}(r)}{dr}) +4r^3\frac{dj(r)}{dr}\overline{\omega}(r) = 0 
\end{equation} 
where
$j(r) = e^{-(\Phi+\Lambda)} = e^{-\Phi(r)}\sqrt{1-\gamma(r)}$
and $\gamma(r) = 1-\frac{2m(r)}{r}$.
Equation (6) is to be integrated from the center of the neutron star towards the surface under the boundary conditions that $\overline{\omega}$ must be regular at the center where r = 0 and $\frac{\overline{d\omega}}{dr}$ vanishes at r = 0. For numerical calculations an arbitrary value of $\omega$ is selected at the center of the neutron star and integration takes place from there, towards the surface. Outside the neutron star $\overline{\omega}$ has the behavior
$\overline{\omega}(r,\Omega) = \Omega - \frac{2}{r^3}J(\Omega)$
where $\Omega$ is the rotation of the neutron star and J($\Omega$) is the angular momentum of the neutron star. The angular momentum can also be given by 
$J(\Omega) = \frac{R^4}{6}(\frac{d\overline{\omega}}{dr})_R.$
The quadrupole equations given in [16] are solved for a rotational frequency 500Hz. The tidal deformability equations given in [17] are essential to determine the shape and deformation of the neutron star caused by rotation about its axis and inspiral rotation in a binary system, respectively.
\section{Results} 
Fig. 1(a) represents the pressure and energy relation, for all three chosen EoSs. The strange matter EoS (MIT - black dot) is unphysical in the low energy density region (at the crust). Here, a key point must be stated, i.e. in the low energy density region, there is only nuclear matter. Strange matter or hyperons only come into existence when the energy density increases beyond a few times the nuclear matter density. All of them have only nuclear matter at low energy density, so they must show a similar characteristic in that region. As the energy density increases, around 580 MeV/fm$^3$, F-QMC700 starts to deviate from the N-QMC700 because hyperons ($\Xi^-,\Lambda,\Xi^0$) start to develop at this point (see Ref.[7]).  On the other hand, the strange matter EoS  around the energy density 300 MeV/fm$^3$ and 500 MeV/fm$^3$ crosses the nucleons only matter EoS, suggesting a transition to strange quark matter (deconfined quarks) often called a "strange star".  N-QMC700 is the stiffest EoS which is based on nucleon only matter. Fig. 1(b) gives the relation between the mass and the radius of neutron stars for these EoSs. Although the strange matter EoS predicts a maximum mass of over 2 $M\odot$, it poorly predicts the mass and the radii of the neutron stars at the low energy densities. At the crust, where the energy density is low, it must show an expansion in the radius, like the N-QMC700 and F-QMC700.
\begin{figure}[h!]
	\subfloat[]{\includegraphics[height=5.1cm,width=7.5cm]{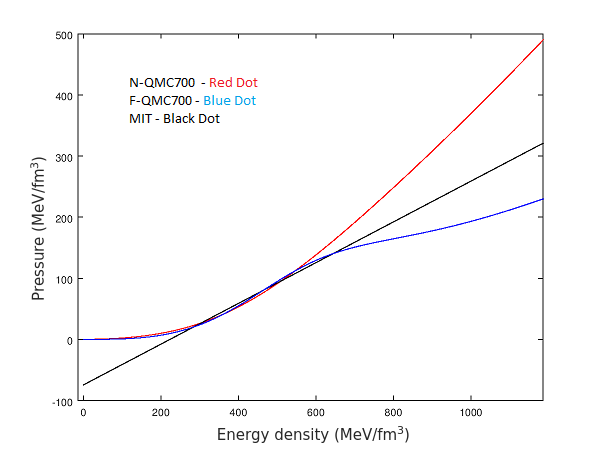} }%
	\qquad
	\subfloat[]{\includegraphics[height=5.1cm,width=7.5cm]{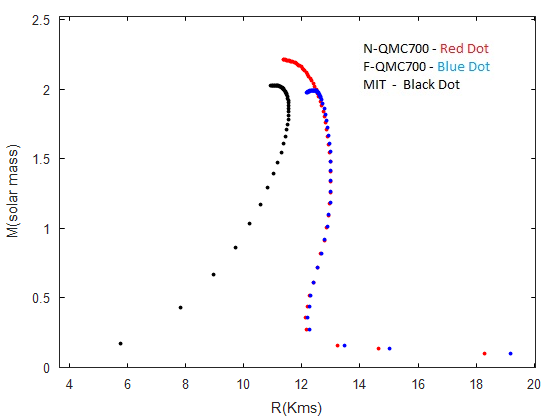} }%
	\caption{(a) Pressure-Energy density plot for selected EoSs, (b)  Mass-Radius plot for selected EoSs.}
\end{figure}
\begin{figure}[h!]
	\begin{center}
		\subfloat[]{\includegraphics[height=5.0cm,width=7.1cm]{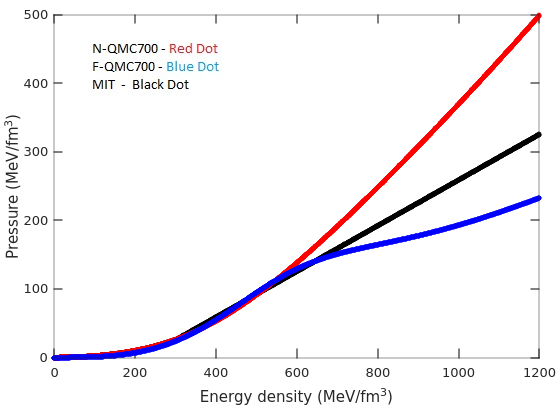} }%
		\qquad
		\subfloat[]{\includegraphics[height=5.0cm,width=7.5cm]{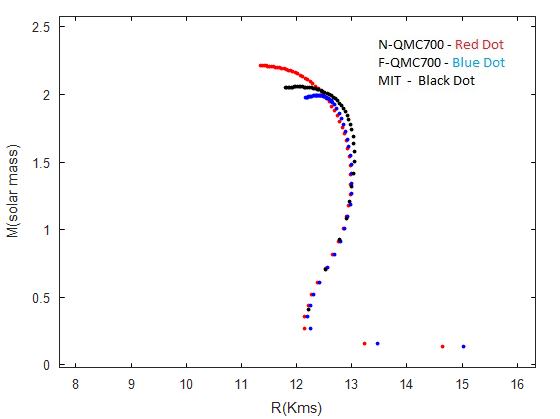} }%
		\caption{(a) Pressure-Energy density plot for selected EoSs with strange matter EoS parameters changed to N-QMC700 EoS parameters for the lower energy densities below 300 MeV/fm$^3$, (b) Mass-Radius plot for EoSs given in Fig. 2(a)}%
	\end{center}
\end{figure}
\par In Fig. 2, the strange matter EoS is corrected for the low energy density region with the parameters of nucleons only EoS (N-QMC700). In Fig. 2(a), the strange matter EoS parameters are matched to the nucleon only EoS parameters for the energy densities below 300 MeV/fm$^3$ and in Fig. 2(b), mass vs radius plot is presented with changed parameters for strange matter EoS. Here, we are calculating the properties of neutron stars when the strange matter starts to exist at 300 MeV/fm$^3$. In Fig. 2, all three EoSs now show the same characteristic at the low energy densities. With this change the behavior of all the three EoS show similar radii at the low energy densities and all three EoSs predict maximum mass of neutron stars close to 2$M_\odot$. There is a universality that most of the EoSs dimensionless moment of inertia vs dimensionless quadrupole moment (I-Q relation) plots fall  on the top of each other and show a characteristic which is (almost) independent of EoSs [18]. In Fig. 3(a), N-QMC700 and F-QMC700 both show the same characteristic and follow the universality while strange matter EoS behavior is very different and it does not follow the universality. After fixing the parameters of the strange matter EoS in the lower energy density region, it follows the universality depicted in Fig. 3(b). 
As predicted by Yagi and Yunes [17], dimensionless tidal deformability ($\lambda/M^5$) vs dimensionless moment of inertia ($I/M^3$) plot shows a behavior which is independent of EoS. As shown in Fig. 4(a), both N-QMC700 and F-QMC700 EoSs fall exactly on top of each other and follow the universality whereas the strange matter EoS is very different, falling far away from them and not following the universality. The results when the parameters of the strange matter EoS are fixed for the energy density below 300  MeV/fm$^3$ with N-QMC700 are shown in Fig. 4(b). The strange matter EoS dimensionless moment of inertia vs dimensionless tidal deformability plot follows the universality like the QMC EoS. Lattimer and Schultz [19] showed that the parameter of moment of inertia, I/M$_0$R$_0^2$,  where $M_0$ is the mass and $R_0$ is the radius of the neutron star, is related to compactness in such a way that their plot is independent of EoSs and all of them fall almost on the same line. We see in Fig. 5(a) that the moment of inertia parameter for the strange matter EoS is very different from N-QMC700 and F-QMC700 and it does not follow the moment of inertia constraint, while N-QMC700 and F-QMC700 EoS do. After fixing the parameters as shown in Fig. 5(b) at the lower energy densities 300 MeV/fm$^3$, all three EoSs satisfy the relation independent of EoS. They fall almost on the same line [10]. 
\begin{figure}[h!]
	\begin{center}
		\subfloat[]{\includegraphics[height=5.0cm,width=7.2cm]{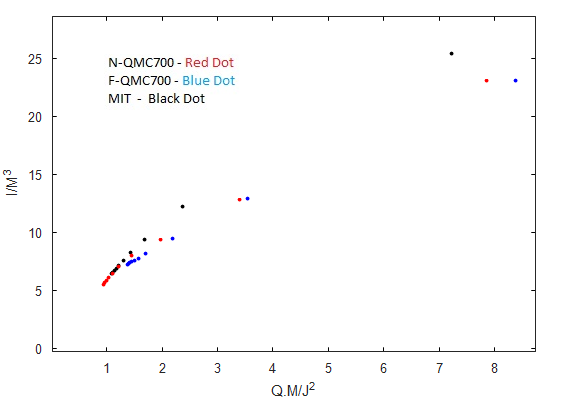} }%
		\qquad
		\subfloat[]{\includegraphics[height=5.0cm,width=7.2cm]{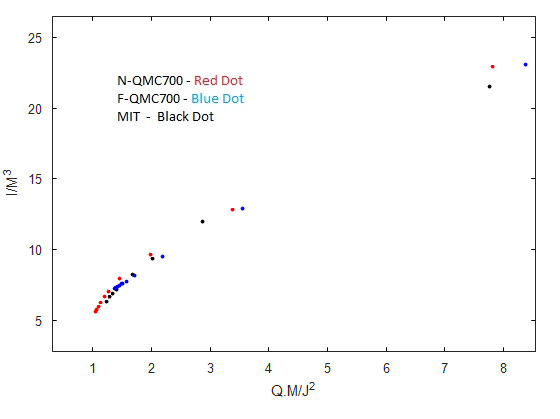} }%
		\caption{Dimensionless moment of inertia vs dimentionless quadrupole moment plot for EoSs given, (a) in Fig 1(a) and (b) in Fig. 2(a).}
	\end{center}
\end{figure}
\begin{figure}[h]
	\begin{center}
		\subfloat[]{\includegraphics[height=5.0cm,width=7.2cm]{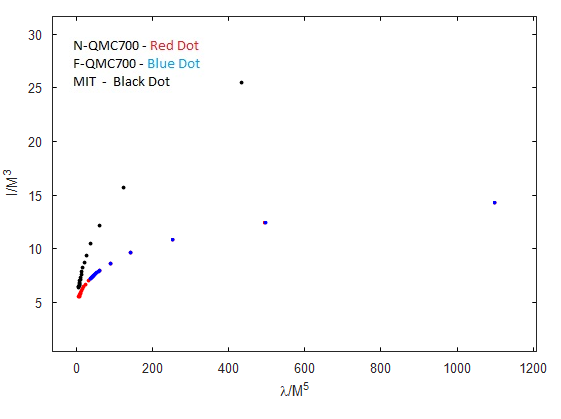} }%
		\qquad
		\subfloat[]{\includegraphics[height=5.0cm,width=7.2cm]{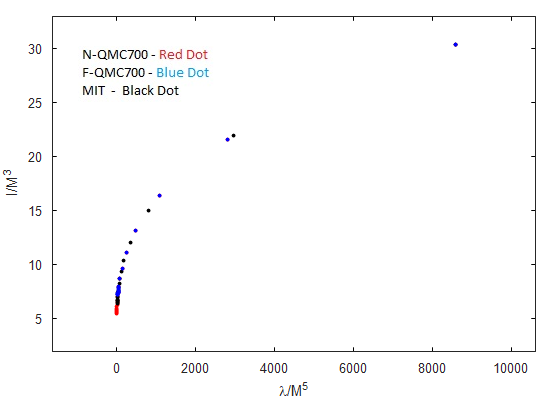} }%
		\caption{Dimensionless tidal deformability vs dimensionless moment of inertia plot for EoSs given, (a) in Fig 1(a) and (b) in Fig. 2(a).}
	\end{center}
\end{figure}
\begin{figure}[H]
	\begin{center}
		\subfloat[]{\includegraphics[height=5.0cm,width=7.2cm]{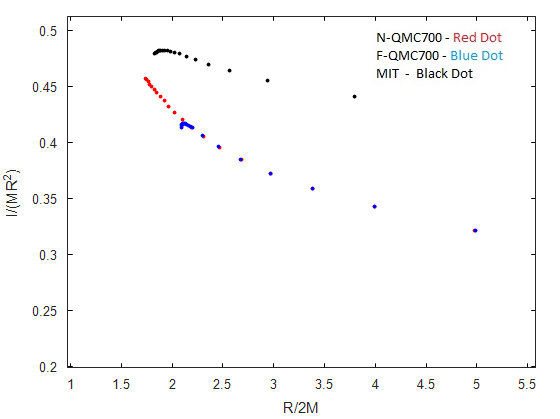} }%
		\qquad
		\subfloat[]{\includegraphics[height=5.0cm,width=7.2cm]{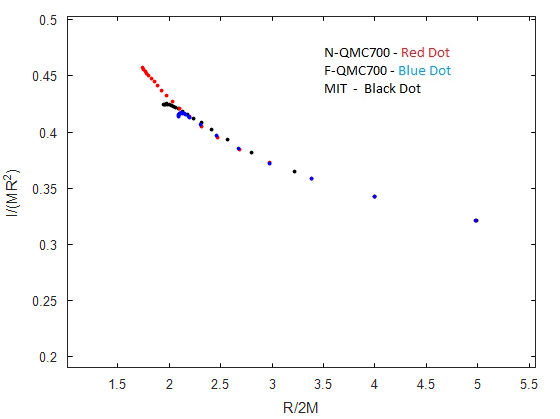} }%
		\caption{Moment of inertia parameter vs R/2M plot for EoSs given, (a) in Fig.1(a) and (b) in Fig. 2(a).}
	\end{center}
\end{figure}
\section{Conclusion and Future Perspectives}
Models for rotating neutron stars have been successfully compared within the Hartle framework at frequency 500 Hz for the different EoSs under the constraints imposed by observations. It is shown that EoSs based purely on hadrons (N-QMC700 \& F-QMC700), satisfy the empherical constraints which is a clear indication that the QMC model may play a significant role in defining the properties of matter at  energy densities several times the  nuclear matter density. Several universalities among the neutron star properties such as dimensionless moment of inertia vs dimensionless quadrupole momentum, dimensionless moment of inertia vs dimensionless tidal deformability and moment of inertia parameter vs R/2M, have been confirmed in these models. \par
Following the discovery of gravitational waves we may impose  new constraints on the EoS in a neutron star binary system, which may prove handy in understanding the interior of neutron stars. Furthermore, recently improvements have been made in the QMC model [20][21], and with these advancements we can improve our EoS and can get better results. In future it will be interesting to see how QMC EoS explain the neutron star cooling process, magnetic fields and the tidal deformability in the neutron star merger. 
\section*{Acknowledgments}
This research is supported by Adelaide Scholarship International, without the aid this study could not have been possible.
\section{References}

\end{document}